\documentclass[letterpaper, 10 pt, conference]{ieeeconf}  % Comment this line out
\IEEEoverridecommandlockouts                              % This command is only
\usepackage{amsmath}
\usepackage{tikz}
\usetikzlibrary{shapes,arrows}
\usepackage{tikz-network}
\usetikzlibrary{positioning}
\usepackage{subcaption}
\usepackage{mathrsfs}
\usepackage{booktabs}

\usepackage{amsthm}

\newtheorem{thm}{Theorem}

\newtheorem{lmm}{Lemma}

\newtheorem{asm}{Assumption}

\newcommand{\thistheoremname}{}
\newtheorem*{genericthm*}{\thistheoremname}
\newenvironment{namedthm*}[1]
  {\renewcommand{\thistheoremname}{#1}%
  \begin{genericthm*}}
  {\end{genericthm*}}

\title{\LARGE \bf
Robust Queue Length Estimation for Ramp Metering \\
in a Connected Vehicle Environment
}

\author{Yu Tang, Kaan Ozbay and Li Jin
\thanks{This work was in part supported by US NSF Award CMMI-1949710, USDOT Award \# 69A3551747124, NSFC Project 62103260 via the C2SMART Center, NYU Tandon School of Engineering, SJTU UM Joint Institute, and J. Wu \& J. Sun Endowment Fund.}% <-this % stops a space
\thanks{Y. Tang is with C2SMART Center, Department of Civil \& Urban engineering, Tandon School of Engineering, New York University, 11201, USA.
        {\tt\small tangyu@nyu.edu}}%
\thanks{K. Ozbay is with C2SMART Center, Department of Civil \& Urban engineering, Tandon School of Engineering, New York University, 11201, USA.
        {\tt\small kaan.ozbay@nyu.edu}}
\thanks{L. Jin is with UM Joint Institute and Department of Automation, Shanghai Jiao Tong University, Shanghai, 200240, China.
        {\tt\small li.jin@sjtu.edu.cn}}%
}

\begin{document}

\maketitle
\thispagestyle{empty}
\pagestyle{empty}

%%%%%%%%%%%%%%%%%%%%%%%%%%%%%%%%%%%%%%%%%%%%%%%%%%%%%%%%%%%%%%%%%%%%%%%%%%%%%%%%
\begin{abstract}
Connected vehicles (CVs) can provide numerous new data via vehicle-to-vehicle or vehicle-to-infrastructure communication. These data can in turn be used to facilitate real-time traffic state estimation. In this paper, we focus on ramp queue length estimation in a connected vehicle environment, which improves control design and implementation of ramp metering algorithms. One major challenge of the estimation problem is that the CV data only represent partial traffic observations and could introduce new uncertainties if real-time CV penetration rates are unknown. To address this, we build our estimation approach on both traditional freeway sensors and new CV data. We first formulate a ramp queue model that considers i) variations in the penetration rate and ii) noise in measurements. Then we develop a robust filter that minimizes the impacts of these two kinds of uncertainties on queue estimation. More importantly, we show that the designed filter has guaranteed long-term estimation accuracy. It allows us to quantify in a theoretical way the relationship between estimation error and fluctuation of CV penetration rates. We also provide a series of simulation results to verify our approach.

%Ramp metering plays an important part in freeway control that addresses traffic congestion. It has been pointed out that this strategy can achieve better control performance if ramp queue length is available to controllers. However, typical traffic sensors, such as induction loop sensors, cannot measure the queue length. In this paper, we consider ramp queue estimation in a connected vehicle environment since connected vehicles can provide numerous new data, e.g. during vehicle-to-infrastructure communication, which makes the estimation easier. We first.
\end{abstract}

\section{INTRODUCTION}

%\subsection{Motivation}

Ramp metering is one of the major freeway management strategies that can be used to alleviate both recurrent and non-recurrent traffic congestion.
It typically regulates on-ramp flows to improve mainline mobility, at the same time keeping reasonable queue lengths at ramps \cite{papageorgiou2002freeway,liu2021deep,tang2021resilient}. 
Field evaluations have demonstrated that well-designed ramp metering can significantly improve traffic efficiency \cite{papageorgiou1997alinea, bhouri2013isolated,wu2018field}. In current practice, ramp metering is implemented mainly with the aid of induction loop detectors so that ramp controllers can adjusted metered flows, based on real-time traffic volumes, speed and occupancy \cite{grzybowska2022ramp}.

It is also reported that the current control of ramp flows can be refined, with smoothing metering rates and reducing traffic oscillation \cite{gordon1996algorithm, sun2005localized,papamichail2008traffic}, if on-ramp queue length is available to controllers. But the ramp queue length is not easily acquired from widely-used induction loop detectors; it can only be estimated from these sensors rather than be measured directly. The estimation methods are usually built on deliberate assumptions of vehicle characteristics and sensor locations
\cite{sun2005localized,papamichail2010heuristic,vigos2006ramp,wu2008methodologies}, and their applicability could be restricted consequently.

The estimation problem could become more tractable, thanks to the emergence of connected vehicles (CVs). In a connected vehicle environment, vehicle-to-vehicle or vehicle-to-infrastructure communication yields substantial new data, such as basic safety messages (BSMs), that contain detailed information including vehicle position, speed, acceleration and so on \cite{liu2016delivering}. Though these data are only from partial observations of freeway traffic (mainly related to CVs), it is still possible to convert them into traffic measurements \cite{chen2021generating,vasudevan2022algorithms} and enhance freeway state estimation \cite{bekiaris2016highway, khan2017real, fountoulakis2017highway, seo2017traffic, zhao2021generic, wang2022real}.

\begin{figure}[htbp]
    \centering
    \includegraphics[width=\linewidth]{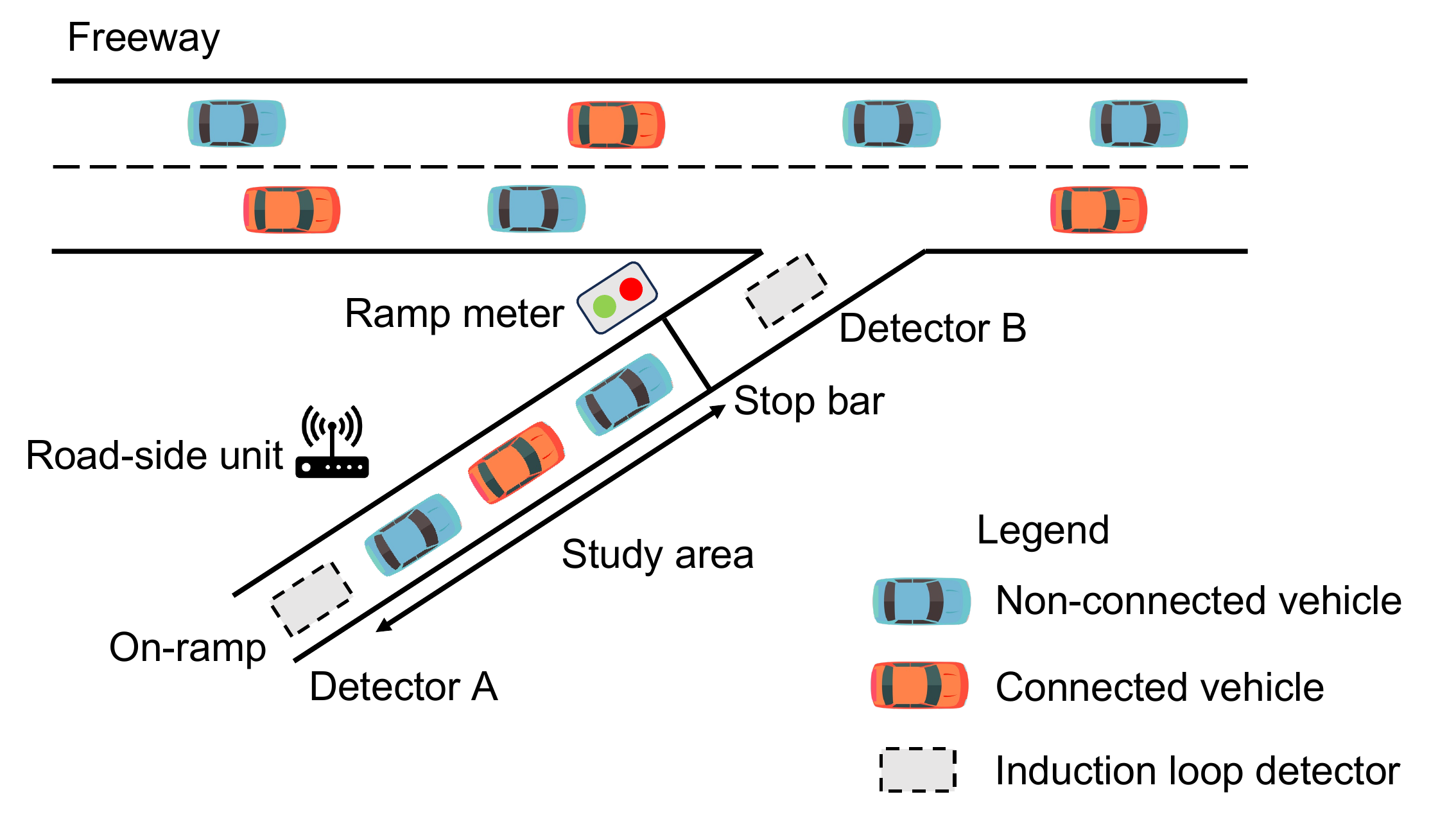}
    \caption{Problem setting: our target is to assess ramp queue length based on induction loop detectors and Basic Safety Messages (BSMs) \cite{vasudevan2022algorithms}. We assume that induction loop detectors are deployed at both ends of the on-ramp, which is a minimum requirement by ramp metering \cite{adot2013ramp};  we also assume that road-side units are deployed to collect BSMs. Note that BSMs contain information of vehicle positions, and thus it is feasible to recognize those from CVs at the on-ramp and to extract traffic data of the corresponding CVs   \cite{chen2021generating,vasudevan2022algorithms}.
    }
    \label{fig_modeling}
\end{figure}

In this paper, we aim at developing a robust queue length estimation approach that utilizes both traditional sensors (e.g. induction loop detectors) and emerging CV data (e.g. BSMs); see Fig.~\ref{fig_modeling}. It should be pointed out that queue length at a certain link may refer to i) distance from the first stopping vehicle to the last one \cite{cheng2011cycle,yin2018kalman} or ii) simply link-level vehicle number. The latter suffices for ramp metering \cite{vigos2006ramp,wu2008methodologies,papamichail2010heuristic,sun2005localized}, and following this convention, we focus on  estimating the vehicle number \footnote{We use the terms queue length/vehicle number interchangeably in this paper.} between the stop bar and the on-ramp entrance as shown in Fig.~\ref{fig_modeling}. To this end, we first formulate a ramp queue model, a dynamic system that considers the fluctuating penetration rate of CVs and noisy traffic measurements. Particularly, the variation in the penetration rate induces model parameter uncertainty. Then we propose a robust filter, based on the queue model, that produces the estimation with performance guarantees, even under the parameter uncertainty and the measurement noise.

Previous studies have adopted Kalman filter for queue length estimation, which mainly deals with measurement noise. Given only induction loop detectors, Kalman filters were built on the assumptions of uniform vehicle length and headway, internal sensors in the middle of on-ramps, and so on \cite{vigos2006ramp,wu2008methodologies,papamichail2010heuristic}. Though numerical examples have demonstrated that those filters could still work given heterogeneous vehicles, e.g. with various length, the internal sensors are indispensable. Such sensors, however, may be unavailable in practice \cite{adot2013ramp}, impeding application of those estimation approaches. Kalman filters were also developed in a connected vehicle environment \cite{yin2018kalman,wang2022real}. These methods ignored the uncertainty of CV penetration rates, e.g. assuming the rate estimated from other approaches \cite{wang2022real}, and thus failed to quantify in a theoretical way how uncertain CV penetrate rates influence estimation accuracy. To our best knowledge, limited work has discussed this problem.

Our major contribution is to introduce a robust filter with a performance guarantee (Theorem~\ref{thm_1}). Here the robustness is two-fold: the filter minimizes the impacts of both measurement noise and uncertain CV penetration rates on queue length estimation. Moreover, Theorem~\ref{thm_1} allows us to quantify the relationship between estimation accuracy and the uncertain penetration rates. We also demonstrate these via a series of simulation results. It should be pointed out that our approach is related to robust $H_\infty$ filters. Though the general theories of $H_\infty$ filters have been well developed \cite{chang2014robust}, it is unclear how to apply them to queue length estimation. As we will see, the typical design of $H_\infty$ filters does not work in our problem; we need to consider one of its extensions \cite{nguyen2018robust} and develop the filter based on our ramp queue model.

The rest of the paper is organized as follows. Section~\ref{sec_modeling} introduces the ramp queue model.
Section~\ref{sec_stability} presents our robust filter design. Section~\ref{sec_example} provides a case study. Finally, Section~\ref{sec_conclusion} summarizes our work and discusses future research.

%%%%%%%%%%%%%%%%%%%%%%%%%%%%%%%%%%%%%%%%%%%%%%%%%%%%%%%%%%%%%%%%%%%%%%%%%%%%%%%%

\section{Modeling and formulation}
\label{sec_modeling}

We present the ramp queue model and formally state the estimation problem in this section.

Consider queuing vehicles at the on-ramp as shown in Fig.~\ref{fig_modeling}. We denote by $x_{\mathrm{all}}(t)$ (resp. $x_{\mathrm{cv}}(t)$ ) the number of total vehicles (resp. CVs) between the stop bar and the detector A in Fig.~\ref{fig_modeling}. Besides, we let $f_{\mathrm{all}}^{\mathrm{in}}(t)$ (resp. $f_{\mathrm{all}}^{\mathrm{out}}(t)$) denote the total flow passing the detector A (resp. detector B) in Fig.~\ref{fig_modeling}. Among them, we denote by $f_{\mathrm{cv}}^{\mathrm{in}}(t)$ (resp.  $f_{\mathrm{cv}}^{\mathrm{out}}(t)$) the inflow (resp. outflow) of CVs. Clearly, it follows $x_{\mathrm{all}}(t)\geq x_{\mathrm{cv}}(t)$, $f_{\mathrm{all}}^{\mathrm{in}}(t)\geq f_{\mathrm{cv}}^{\mathrm{in}}(t)$ and $f_{\mathrm{all}}^{\mathrm{out}}(t)\geq f_{\mathrm{cv}}^{\mathrm{out}}(t)$ for $t=0,1,\cdots$. By the conservation law, we have
\begin{equation}
    \begin{bmatrix}
        x_{\mathrm{all}}(t+1) \\
        x_{\mathrm{cv}}(t+1)
    \end{bmatrix}
    = 
    \begin{bmatrix}
        x_{\mathrm{all}}(t) \\
        x_{\mathrm{cv}}(t)
    \end{bmatrix}
    +\delta_t
    \begin{bmatrix}
        f^{\mathrm{in}}_{\mathrm{all}}(t) \\
        f^{\mathrm{in}}_{\mathrm{cv}}(t)
    \end{bmatrix}
    -\delta_t
    \begin{bmatrix}
        f^{\mathrm{out}}_{\mathrm{all}}(t) \\
        f^{\mathrm{out}}_{\mathrm{cv}}(t)
    \end{bmatrix} \label{eq_conservation},
\end{equation}
where $\delta_t$ is the cycle length of ramp metering, e.g. 30 seconds. That is, the queue length is updated after each control cycle. It helps with determining metering rates in the next cycle. 

Though we do not explicitly model the relation between the vehicle numbers and the flows (concrete models are available in \cite{ferrara2018freeway}), we consider the inflows and outflows subject to the following constraints
\begin{subequations}
    \begin{align}
        0\leq& x_{\mathrm{all}}(t) + \delta_t(f_{\mathrm{all}}^{\mathrm{in}}(t) - f_{\mathrm{all}}^{\mathrm{out}}(t)) \leq Q, t=0,1,\cdots, \\
        0\leq& x_{\mathrm{cv}}(t) + \delta_t(f_{\mathrm{cv}}^{\mathrm{in}}(t) - f_{\mathrm{cv}}^{\mathrm{out}}(t)) \leq Q, t=0,1,\cdots,        
    \end{align}
\end{subequations}
due to limited ramp space, where $Q$ is the maximum queue length. It indicates $x_{\mathrm{all}}(t), x_{\mathrm{cv}}(t) \in [0, Q]$.  For the sake of convenience,
we also require the CV flows $f_{\mathrm{cv}}^{\mathrm{in}}(t)$ and $f_{\mathrm{cv}}^{\mathrm{out}}(t)$ for $t=0,1,\cdots$ such that
\begin{equation}
    \sum_{t=1}^T x_{\mathrm{cv}}(t) \to \infty, \text{ as } T\to\infty. \label{eq_xcv_inf}
\end{equation}
If \eqref{eq_xcv_inf} does not hold, there exists $T_0>0$ such that $x_{\mathrm{cv}}(t)=0$ for all $t\geq T_0$. It means that there are none of queuing CVs at the on-ramp after time $T_0$, and our method will not apply any more in that case. Note that \eqref{eq_xcv_inf} implies
\begin{equation}
    \sum_{t=1}^T x_{\mathrm{all}}(t) \to \infty, \text{ as } T\to\infty. \label{eq_xall_inf}
\end{equation}

Obviously, there exists the following relationship between $x_{\mathrm{all}}(t)$ and $x_{\mathrm{cv}}(t)$:
\begin{equation}
    x_{\mathrm{cv}}(t) = (\alpha + \theta(t)) x_{\mathrm{all}}(t), \label{eq_xall_xcv}
\end{equation}
where $\alpha+\theta(t)$ denotes the penetration rate of CVs at time $t$. We use $\alpha$ to represent the market penetration rate and $\theta(t)$ to denote the variation at the on-ramp. Besides, we make the assumption below.
\begin{asm} \label{ams_1}
    The market penetration rate $\alpha\in(0,1]$ is known, but the fluctuation $\theta(t)$ is unknown and bounded by
\begin{equation}
    ||\theta(t)||_{\ell_\infty} \leq \Theta \in [0,1], \label{eq_Theta}
\end{equation}
where $||\theta(t)||_{\ell_\infty}:=\sup_{t\geq0} |\theta(t)|$.
\end{asm}
Note that the upper bound $\Theta$ can be roughly calibrated when we have access to measurements of CV flows and total flows;  see \eqref{eq_obs} below. 

We consider the traffic data given by
\begin{equation}
    \begin{bmatrix}
        \tilde{f}_{\mathrm{all}}^{\mathrm{in}}(t) \\
        \tilde{f}_{\mathrm{all}}^{\mathrm{out}}(t)
        \\
        \tilde{f}_{\mathrm{cv}}^{\mathrm{in}}(t) \\
        \tilde{f}_{\mathrm{cv}}^{\mathrm{out}}(t) \\
        \tilde{x}_{\mathrm{cv}}(t)    
    \end{bmatrix} 
    = 
    \begin{bmatrix}
        f_{\mathrm{all}}^{\mathrm{in}}(t) \\
        f_{\mathrm{all}}^{\mathrm{out}}(t)
        \\
        f_{\mathrm{cv}}^{\mathrm{in}}(t) \\
        f_{\mathrm{cv}}^{\mathrm{out}}(t) \\
        x_{\mathrm{cv}}(t)    
    \end{bmatrix} + w(t),  \label{eq_obs}
\end{equation}
where $w(t)\in\mathbb{R}^5$ denotes unknown measurement noise, $\tilde{f}_{\mathrm{all}}^{\mathrm{in}}(t)$ and  $\tilde{f}_{\mathrm{all}}^{\mathrm{out}}(t)$ are measured by induction loop detectors,  $\tilde{f}_{\mathrm{cv}}^{\mathrm{in}}(t)$, $\tilde{f}_{\mathrm{cv}}^{\mathrm{out}}(t)$, $\tilde{x}_{\mathrm{cv}}(t)$ are extracted  from BSMs \cite{chen2021generating}. We only know the measured values due to the noise $w(t)$, and we assume $w(t)$ to satisfy:
\begin{asm} \label{asm_noise}
    The noise $w(t)$ satisfies $||w||_{\ell_2}<\infty$, where
\begin{equation}
||w||_{\ell_2}:=\Big(\sum_{t=0}^\infty ||w(t)||_2^2\Big)^{1/2} = \Big(\sum_{t=0}^\infty w^{\mathrm{T}}(t)w(t)\Big)^{1/2}. \label{eq_noise_asm}
\end{equation}
\end{asm}
Note that \eqref{eq_noise_asm} implies that the noise has finite energy. It is a common technique for modeling  temporary noise \cite{chang2014robust}. 
%Indeed, not every noise satisfies Assumption~\ref{asm_noise}. But as we will see later, our estimation method could still work properly for the practical purpose, even when the noise does not satisfy Assumption~\ref{asm_noise}.

Letting
\begin{equation*}
    x(t) := 
    \begin{bmatrix}
    x_{\mathrm{all}}(t) \\ x_{\mathrm{cv}}(t)    
    \end{bmatrix} \text{ and }
    \tilde{f}(t) := 
    \begin{bmatrix}
        \tilde{f}_{\mathrm{all}}^{\mathrm{in}}(t) \\ 
        \tilde{f}_{\mathrm{all}}^{\mathrm{out}}(t) \\ 
        \tilde{f}_{\mathrm{cv}}^{\mathrm{in}}(t) \\ 
        \tilde{f}_{\mathrm{cv}}^{\mathrm{out}}(t)
    \end{bmatrix},
\end{equation*}
we reformulate \eqref{eq_conservation}, \eqref{eq_xall_xcv} and \eqref{eq_obs} into a dynamic system as follows:
\begin{subequations}
    \begin{align}
        x(t+1) =&(A+\Delta A(t))x(t) + B\tilde{f}(t) + Dw(t), \label{eq_queue_1} \\
        y(t) =& Cx(t) + Ew(t) \label{eq_queue_2}
    \end{align}
\end{subequations}
with an unknown initial condition $x(0)\in[0,Q]^2$, where 
\begin{equation*}
    A = \begin{bmatrix}
        1 & 0 \\ \alpha & 0
    \end{bmatrix}, \Delta A(t)= \begin{bmatrix}
        0 & 0 \\ \theta(t) & 0
    \end{bmatrix},
\end{equation*}
\begin{equation*}
    B=-D = \begin{bmatrix}
        \delta_t & -\delta_t & 0 & 0 & 0 \\
        0 & 0 & \delta_t & -\delta_t & 0
    \end{bmatrix},
\end{equation*}
\begin{equation*}
    C = \begin{bmatrix}
        0 & 1
    \end{bmatrix}, E = \begin{bmatrix}
        0 & 0 & 0 & 0 & 1
    \end{bmatrix}.
\end{equation*}
Here $x(t)$ denotes the model state, $y(t)$ denotes the measured output, and $\Delta A(t)$ represents the parameter uncertainty. Since the observed flows $\tilde{f}(t)$ do not equal to the real flows $f_{\mathrm{all}}^{\mathrm{in}}(t)$, $f_{\mathrm{all}}^{\mathrm{out}}(t)$, $f_{\mathrm{cv}}^{\mathrm{in}}(t)$ and $f_{\mathrm{cv}}^{\mathrm{out}}(t)$, it is necessary to include the measurement noise in the state equation \eqref{eq_queue_1}.

Given the queuing system \eqref{eq_queue_1}-\eqref{eq_queue_2}, our goal is to design a robust filter that estimates the state $x(t)$ given $y(t)$ and $\tilde{f}(t)$ but without knowing the initial condition $x(0)$, the noise $w(t)$ and the parameter uncertainty $\Delta A(t)$.
\section{Filter Design}
\label{sec_stability}

We introduce the proposed filter design in this section. Consider a filter given by
\begin{subequations}
    \begin{align}
    \hat{x}(t+1) =& A\hat{x}(t) + B\tilde{f}(t) + L(y(t)-\hat{y}(t)), \label{eq_observer_1} \\
    \hat{y}(t) =& C\hat{x}(t), \label{eq_observer_2}
\end{align}    
\end{subequations}
with an initial condition $\hat{x}(0)\in[0,Q]^2$, where $L$ is the gain to be designed. Note that we can manually specify the initial condition $\hat{x}(0)$.

Define the estimation error
\begin{equation}
    e(t) = x(t) - \hat{x}(t). \label{eq_error_dfn}
\end{equation}
Plugging \eqref{eq_queue_1}-\eqref{eq_queue_2} and \eqref{eq_observer_1}-\eqref{eq_observer_2} into \eqref{eq_error_dfn} gives
\begin{equation}
    e(t+1) = (A-LC)e(t)+\Delta A(t) x(t) + (D-LE)w(t). \label{eq_errordynamics}
\end{equation}
We aim at designing the filter gain $L$ so that the estimation error $e(t)$ satisfies the following criterion \cite{nguyen2018robust}. That is, there exists $\mu_1,\mu_2>0$ and some positive definite function $\gamma:[0,Q]^2\to\mathbb{R}_{\geq0}$ such that
\begin{align}
        \sum_{t=0}^T ||e(t)||_2^2  \leq& \mu_1 \sum_{t=0}^T ||x(t)||_2^2 
        + \mu_2\sum_{t=0}^T ||w(t)||_2^2 \nonumber \\
        & + \gamma(x(0), \hat{x}(0)), ~T=0,1,\cdots, \label{eq_criterion}
\end{align}
given any initial condition $x(0)$ and $\hat{x}(0)$, the input $u(t)$, the noise $w(t)$ and the parameter uncertainty $\Delta A(t)$. 

The criterion \eqref{eq_criterion} is an extension of the typical objective of designing $H_\infty$ filters \cite{chang2014robust}. We consider it because it is not easy to implement the conventional design of $H_\infty$ filters in our problem. The major difficult is that we need to additionally assume $\tilde{f}(t)=0$ when $x(t)=0$, which unreasonably indicates that the inflows equal zero when there are no queuing vehicles at the on-ramp.

To understand the criterion \eqref{eq_criterion}, we divide the both sides by $\sum_{t=0}^T ||x(t)||_2^2$ and obtain
\begin{equation}
    \frac{\sum_{t=0}^T ||e(t)||_2^2}{\sum_{t=0}^T ||x(t)||_2^2}  \leq \mu_1 
        + \mu_2\frac{\sum_{t=0}^T ||w(t)||_2^2}{{\sum_{t=0}^T ||x(t)||_2^2}} + \frac{\gamma(x(0), \hat{x}(0))}{{\sum_{t=0}^T ||x(t)||_2^2}}. \label{eq_c1}
\end{equation}
Noting i) $||w||_{\ell^2}^2<\infty$ by Assumption~\ref{asm_noise}, ii) $\gamma(x(0),\hat{x}(0))<\infty$ owing to $x(0),\hat{x}(0)\in [0,Q]^2$ and iii) $||x||_{\ell^2}^2=\infty$ by \eqref{eq_xall_inf}, we let $T\to\infty$ and reduce \eqref{eq_c1} to
\begin{equation}
    \frac{||e||_{\ell^2}^2}{||x||_{\ell^2}^2}
     \leq \mu_1. \label{eq_guarantee1}
\end{equation}
Thus, if \eqref{eq_criterion} holds, the long-term relative error is bounded by $\mu_1$, and an upper bound of the long-term error rate is given by $\sqrt{\mu_1}$ (or a lower bound of the long-term estimation accuracy is given by $1-\sqrt{\mu_1}$) \cite{nguyen2018robust}. It indicates that we can minimize $\mu_1$ by designing the filter gain $L$. Now, we turn to $\mu_2$. Recalling \eqref{eq_c1}, we can conclude that $\mu_2$ is related to convergence speed: smaller $\mu_2$ indicates faster convergence.

%Indeed, if Assumption~\ref{asm_noise} does not hold, the performance guarantee \eqref{eq_guarantee1} is not available. But the filter satisfying \eqref{eq_criterion}  may still have acceptable performance in practice. To see that, we assume that $w(t)$ is persistent bounded noise, i.e. $\sup_{t\geq0}|w(t)|<\infty$, which is a rather general assumption. In that case, we consider estimation over a finite interval $[0, T_{\max}]$. This may suffice for the practical purpose since ramp metering usually runs only during peak hours \cite{adot2013ramp}. As seen in \eqref{eq_c1}, $\sum_{t=0}^T||w(t)||_2^2$ is suppressed by $\mu_2$, even when $T=T_{\mathrm{max}}$. Moreover, the impacts of $\sum_{t=0}^T||w(t)||_2^2$ and $\gamma(x(0),\hat{x}(0))$ decay as $\sum_{t=0}^T||x(t)||_2^2$ is non-decreasing with respect to $T$. Thus, to control the relative error, we can first set a small value for $\mu_2$ and then minimize $\mu_1$ by optimizing the filter gain $L$.

%we can first set small $\mu_2$ to reduce the impact of  $\sum_{t=0}^{T_{\mathrm{max}}}||w(t)||_2^2$ on and then minimize $\mu_1$. Note that.

For the filter design, we have the following result:
\begin{thm} \label{thm_1}
    Consider the queuing system \eqref{eq_queue_1}-\eqref{eq_queue_2} with the filter \eqref{eq_observer_1}-\eqref{eq_observer_2}. If there exists a symmetric positive definite matrix $P\succ0$, a vector $R\in\mathbb{R}^{2}$ and positive scalars $\mu_1,\mu_2,\mu_3>0$ such that 
    \begin{equation}
    \begin{bmatrix}
    -\mu_3I & \Theta P & 0 & 0 & 0 \\
    * & -P & PA-RC & 0 & PD-RE \\
    * & * & -P+I & 0 & 0  \\
    * & * & * & (\mu_3-\mu_1) I & 0  \\
    * & * & * & * & -\mu_2 I
    \end{bmatrix} \prec 0, \label{eq_thm}
    \end{equation}
where $I$ denotes the identity matrix, then the filter gain is given by
\begin{equation}
    L = P^{-1}R,
\end{equation}
and the filter \eqref{eq_observer_1}-\eqref{eq_observer_2} satisfies \eqref{eq_criterion}.
\end{thm}
The proof of Theorem~\ref{thm_1} is available in Appendix~\ref{app_pf_thm1}. Note that $X\succ 0$ (resp. $X\prec 0$) denotes that $X$ is a positive (resp. negative) definite matrix, and $X\succeq 0$ (resp. $X\preceq 0$) implies that $X$ is a semi-positive (resp. semi-negative) definite matrix. We also use ``$*$'' to simplify notations of symmetric matrices, e.g.
\begin{equation*}
    \begin{bmatrix}
        X_{11} & X_{12} \\ * & X_{22}
    \end{bmatrix} 
    :=\begin{bmatrix}
        X_{11} & X_{12} \\ X^\mathrm{T}_{12} & X_{22}
    \end{bmatrix}. 
\end{equation*}

Recalling the elaboration of \eqref{eq_criterion}, we can first specify $\beta$ and then solve
\begin{equation}
        (\mathrm{P}_1)~\min_{\mu_1,\mu_2,\mu_3>0,  P,R\succ0}~ \mu_1+\beta\mu_2 ~ s.t.~\eqref{eq_thm}.
\end{equation}
Since $\mu_1$ is related to long-term estimation error, we can quantify, by solving $\mathrm{P}_1$, the relationship between the long-term estimation accuracy and the uncertainty bound $\Theta$. Note that $\mathrm{P}_1$ belongs to convex programming, and hence it can be easily solved by Mosek \cite{aps2019mosek}.

\section{Case Study}
\label{sec_example}

We present a real-world case study, based on  a model developed using the microscopic traffic simulation software, SUMO, to illustrate our robust filter design. We first present a sensitivity analysis that reveals impact of the penetration rates of CVs and measurement noise. Then we focus on comparison of various estimation methods. 

\begin{figure}[htbp]
    \centering
    \begin{subfigure}{\linewidth}
    \centering
    \includegraphics[width=\linewidth]{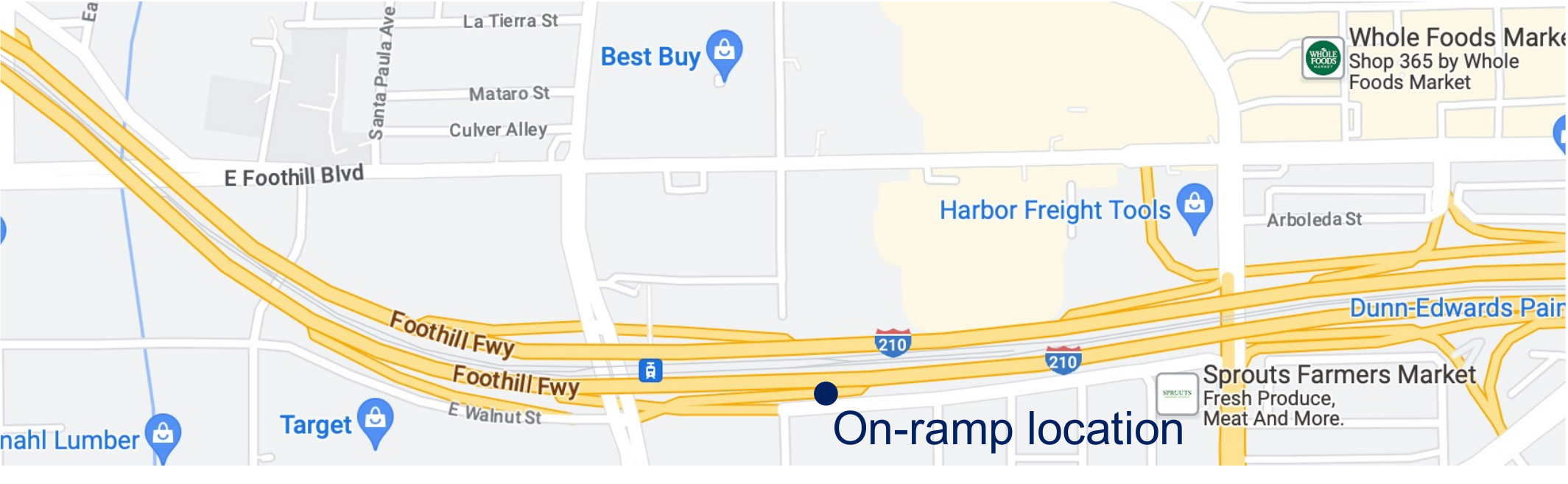}  
  \caption{Selected on-ramp.}
  \label{fig_ramploc}
\end{subfigure}

\begin{subfigure}{\linewidth}
  \centering
  \includegraphics[width=\linewidth]{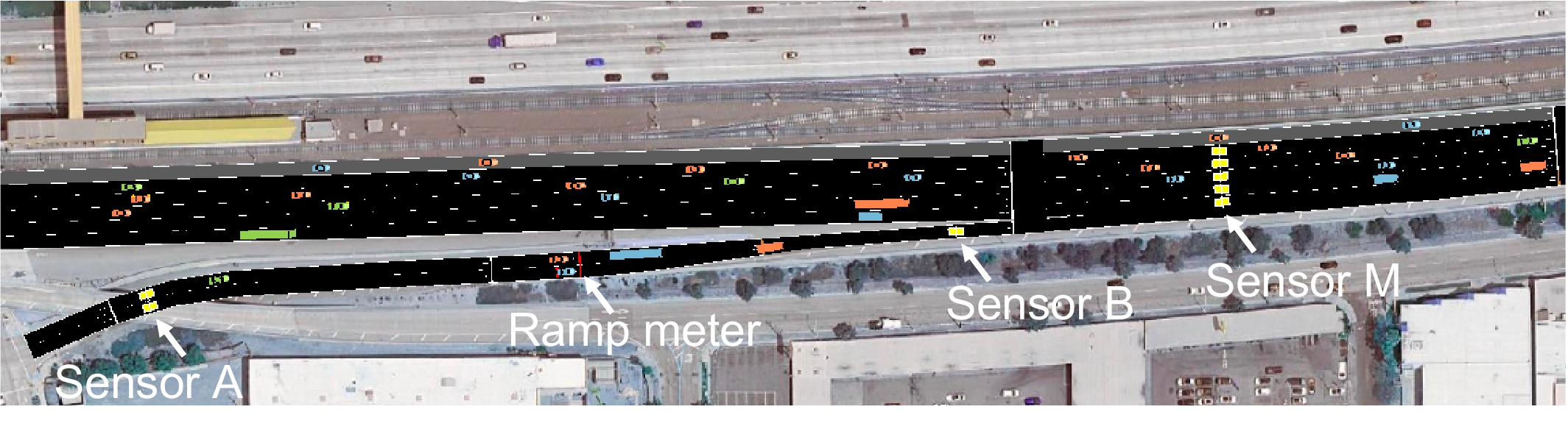}  
  \caption{SUMO setting.}
  \label{fig_sumosetting}
\end{subfigure}
    \caption{Simulation model developed in SUMO.}
    \label{fig_casebackground}
\end{figure}

We consider an on-ramp of Interstate 210 Eastbound (I-210 E) in Los Angeles, California; see Fig.~\ref{fig_ramploc}. The ramp meter, detectors A and B at the on-ramp are set based on the real locations (see Fig.~\ref{fig_sumosetting}), and we have the maximum queue length $Q_{\max}=32$. We also deploy detector M at the mainline to implement the following ramp metering algorithm ALINEA \cite{papageorgiou1997alinea}:
\begin{subequations}
    \begin{align}
        r_i(t) =& \mathrm{mid}\{\underline{R},r_i(t-1) + K_i(\tilde{o}_{\mathrm{M}}-o_{\mathrm{M}}(t)),\bar{R}\} \label{eq_integral} \\
        %r_q(t) =& -\frac{1}{\delta_t}(Q_{\max}-\hat{x}_{\mathrm{all}}(t)) + \tilde{f}_{\mathrm{all}}^{\mathrm{in}}(t-1)  \\
        r_o(t) =& \begin{cases}
            \underline{R} &  \text{if } o_{\mathrm{A}}(t)< \tilde{o}_{\mathrm{A}} \\
            \bar{R} &  \text{if } o_{\mathrm{A}}(t) \geq \tilde{o}_{\mathrm{A}}
        \end{cases} \label{eq_override}\\
        r(t) =& \max\{r_i(t),r_o(t)\} \label{eq_final}
    \end{align}
\end{subequations}
where $r_i(t)$ denotes the metered flow at time $t$ given by integral control with gain $K_i$, $r_o(t)$ denotes the metered flow at time $t$ given by the queue-override strategy, $r(t)$ is the selected metered flow as control input. Briefly speaking, \eqref{eq_integral} determines the metered flow in order to keep the mainline occupancy $o_{\mathrm{M}}(t)$, measured by detector M, close to the critical occupancy $\tilde{o}_{\mathrm{M}}$, and it also clips the metered flow with the minimum (resp. maximum) metered flow $\underline{R}$ (resp. $\bar{R}$) to avoid the well-known wind-up phenomenon \cite{papamichail2008traffic}, where the $\mathrm{mid}$ operator takes the middle value of all the members. \eqref{eq_override}-\eqref{eq_final} indicate that when the occupancy $o_{\mathrm{A}}(t)$ at the ramp entrance, measured by detector A, is higher than the threshold $\tilde{o}_{\mathrm{A}}$, it is necessary to choose the maximum metered flow to discharge queuing vehicles at the on-ramp as quickly as possible. Since we adopt the one-car-per green policy \cite{papageorgiou2002freeway}, we finally convert the metered flow $r(t)$ into red time to realize ramp metering in SUMO. It should be pointed out the algorithm above does not depend on ramp queue length. The reason lies in that we focus on evaluating ramp queue length estimation here, and this algorithm provides the same benchmark when we compare different estimation methods.

\begin{figure}[htbp]
    \centering
    \begin{subfigure}{0.9\linewidth}
    \centering
    \includegraphics[width=\linewidth]{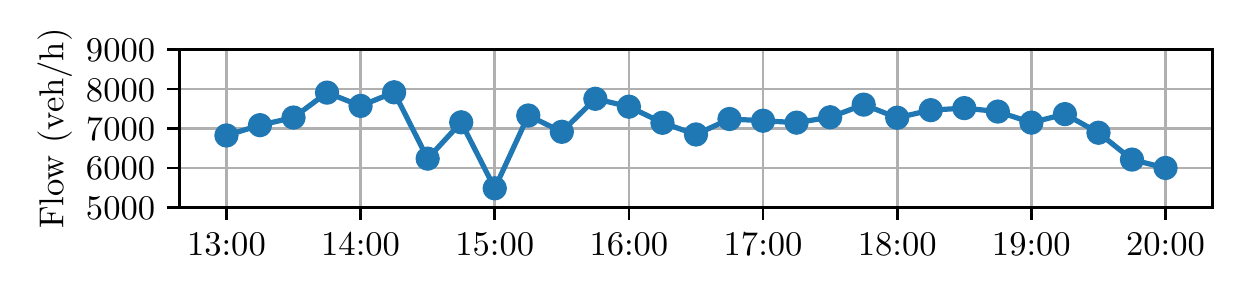}  
  \caption{Upstream mainline demand.}
  \label{fig_mainflow}
\end{subfigure}

\begin{subfigure}{0.9\linewidth}
  \centering
  \includegraphics[width=\linewidth]{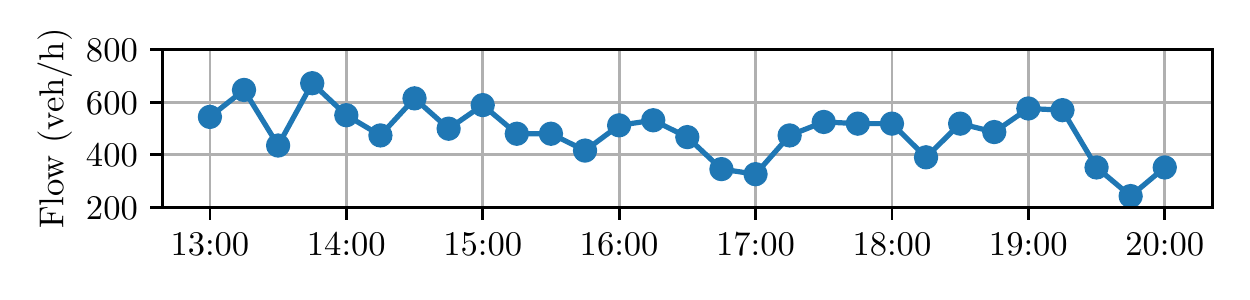}  
  \caption{On-ramp demand.}
  \label{fig_rampflow}
\end{subfigure}
    \caption{Traffic demands.}
    \label{fig_deman}
\end{figure}

The traffic demands considered are illustrated in Fig~\ref{fig_deman}. For vehicle composition, we assume that $95\%$ are passenger vehicles, $3\%$ are light-duty trucks and the remaining are heavy-duty trucks. We consider the period 13:00-20:00 since the peak hours of I-210 E are usually from 14:00-19:00. We use the first half an hour as the warm-up period and start the estimation at 13:30. 

\subsection{Sensitivity analysis}
\begin{figure}[htbp]
    \centering
    \begin{subfigure}{0.49\linewidth}
    \centering
    \includegraphics[width=\linewidth]{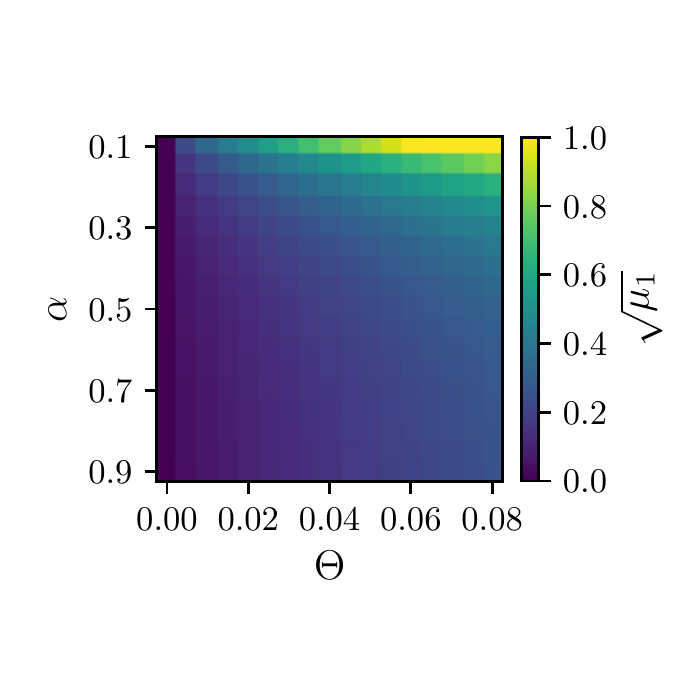}  
  \caption{Impacts on $\mu_1$.}
  \label{fig_sensi_mu1}
\end{subfigure}
\begin{subfigure}{0.49\linewidth}
  \centering
  \includegraphics[width=\linewidth]{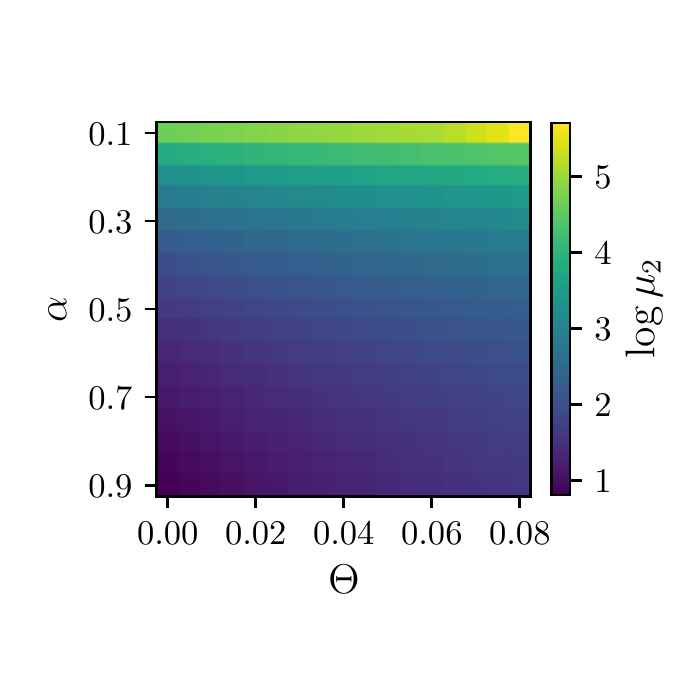}  
  \caption{Impacts on $\mu_2$.}
  \label{fig_sensi_mu2}
\end{subfigure}
    \caption{Sensitivity analysis of market penetration rate of CVs,  $\alpha$ and uncertainty bound $\Theta$.}
    \label{fig_sensitivity}
\end{figure}

The following presents sensitivity analysis of CV penetration rates and measurement noise. We first solve the problem $\mathrm{P}_1$ given different combinations of $\alpha$ and $\Theta$. We choose the weight $\beta=0.1$ since we put priority to the error rate. The results are presented in Fig.~\ref{fig_sensitivity}. We show $\sqrt{\mu_1}$ in Fig.~\eqref{fig_sensi_mu1} since it denotes an upper bound of long-term error rate. It is reasonable to see that when the fluctuation is mild with small $\Theta$, the error upper bound is relatively low. We also present $\log{\mu_2}$ in Fig.~\eqref{fig_sensi_mu2}. We consider $\log{\mu_2}$ since it could be large when $\alpha$ is small but $\Theta$ is large. Recalling that $\mu_2$ is related to convergence speed, we conclude that the convergence could be vary slow, given small $\alpha$ and large $\Theta$, when the measurement noise occurs.

Then, we analyze the filter performance in the simulation. Note that we only specify the CV market penetration rate in advance and do not manipulate the variation bound $\Theta$, which requires us to estimate $\Theta$. It turns out that $\theta(t)$ are relatively scattered given any $\alpha\in\{0.1, 0.3, \cdots, 0.9\}$ and it is not easy to design the filter gain $L$ if we consider the bound $\Theta$ satisfying \eqref{eq_Theta}. Thus we consider $\Theta$ such that $|\theta(t)|\leq\Theta$ for enough $t$. We find that $\Theta=0.08$ suffices for $\alpha\in\{0.1, 0.3, \cdots, 0.9\}$, which is used to derive the filter gain. To test our robust filter, we also assume that the measurement noise happens from 15:00-17:00. Concretely, we assume that $f^{\mathrm{out}}_{\mathrm{cv}}(t)$, $f^{\mathrm{in}}_{\mathrm{cv}}(t)$, $f^{\mathrm{out}}_{\mathrm{all}}(t)$, $f^{\mathrm{out}}_{\mathrm{all}}(t)$ (in unit of veh/h) are subject to bounded noise uniformly distributed on $[-60 ~\mathrm{veh/h}, 60~\mathrm{veh/h}]$ and $x_{\mathrm{cv}}(t)$ (in unit of veh) is subject to bounded noise uniformly distributed on $[-2 ~\mathrm{veh}, 2~\mathrm{veh}]$. The result is summarized in Table~\ref{tab_cv}. We consider the upper bound $\sqrt{\mu_1}$,  $\sqrt{\hat{\mu}_1}$ and the root mean square Error (RMSE), where $\hat{\mu}_1$ is obtained from the simulation:
\begin{equation}
    \hat{\mu}_1 = \sum_{t=0}^{T}||e(t)||_2^2/||x(t)||_2^2.
\end{equation}
Clearly, $\sqrt{\mu_1}$ denotes a theoretical upper bound and $\sqrt{\hat{\mu}_1}$ denotes an empirical evaluation. It is reasonable to find that given higher CV penetration rate, our robust filter yields lower error.
\begin{table}[htbp]
\centering
\footnotesize
\caption{Sensitivity analysis of CV penetration rates.}
\label{tab_cv}
\begin{tabular}{@{}ccccccc@{}}
\toprule
Market penetration rate & 0.1 & 0.3 & 0.5 & 0.7 & 0.9 \\ \midrule
$\sqrt{\mu_1}$ &  1 & 0.4491  & 0.3164 & 0.2712 &     0.2537 \\
$\sqrt{\hat{\mu}_1}$  & 0.6540  & 0.3708 & 0.2701 & 0.1845 & 0.1288   \\
RMSE  & 9.629  & 5.459  & 3.976  & 2.716 & 1.896     \\
\bottomrule
\end{tabular}
\end{table}

We also test our filter given different uniformly bounded noise. We consider the flows $f^{\mathrm{out}}_{\mathrm{cv}}(t)$, $f^{\mathrm{in}}_{\mathrm{cv}}(t)$, $f^{\mathrm{out}}_{\mathrm{all}}(t)$, $f^{\mathrm{out}}_{\mathrm{all}}(t)$ are subjected to bounded noise $[-U, U]$ from 15:00-17:00. The result is given in Table~\ref{tab_noise}. Noting the on-ramp flow shown in Fig.~\ref{fig_rampflow}, we conclude that our filter is robust even when the noise is relatively large, e.g. $U=180$ veh/h.
\begin{table}[hbt]
\centering
\footnotesize
\caption{Sensitivity analysis of measurement noise.}
\label{tab_noise}
\begin{tabular}{@{}ccccccc@{}}
\toprule
Noise bound $U$ (veh/h) & 60 & 90 & 120 & 150 & 180 \\ \midrule
$\sqrt{\hat{\mu}_1}$ ($\alpha=0.1$)  & 0.6540 & 0.6542 & 0.6546 & 0.6550 & 0.6554 \\
$\sqrt{\hat{\mu}_1}$ ($\alpha=0.5$)  & 0.2701 & 0.2708  & 0.2717 & 0.2728 & 0.2742 \\
$\sqrt{\hat{\mu}_1}$ ($\alpha=0.9$)  & 0.1288 & 0.1307 & 0.1330 &  0.1357 & 0.1389 \\
\bottomrule
\end{tabular}
\end{table}

\subsection{Estimation comparison}

We consider another two estimation methods as our baselines. The first one is given by 
\begin{equation}
    \hat{x}_{\mathrm{all}}(t) = \frac{2\tilde{x}_{\mathrm{cv}}(t)}{\tilde{f}_{\mathrm{cv}}^{\mathrm{in}}(t)/\tilde{f}_{\mathrm{all}}^{\mathrm{in}}(t)+\tilde{f}_{\mathrm{cv}}^{\mathrm{out}}(t)/\tilde{f}_{\mathrm{all}}^{\mathrm{out}}(t)}. \label{eq_openloopestimator}
\end{equation}
It evaluates the CV penetration rate at the on-ramp based on the inflows and outflows and converts the observed $\tilde{x}_{\mathrm{cv}}(t)$ to $\hat{x}_{\mathrm{all}}(t)$. The second one was proposed based on the Kalman filter \cite{papamichail2010heuristic}, given by
\begin{align}
    \hat{x}_{\mathrm{all}}(t) =& \hat{x}_{\mathrm{all}}(t-1) + \delta_t(\tilde{f}_{\mathrm{all}}^{\mathrm{in}}(t) - \tilde{f}_{\mathrm{all}}^{\mathrm{out}}(t)) \nonumber \\
    & + K_f(Q_{\mathrm{bb}}\frac{\bar{L}_{\mathrm{veh}}}{\bar{L}_{\mathrm{veh}}+L_{\mathrm{d}}} o_{\mathrm{A}}(t-1)-\hat{x}_{\mathrm{all}}(t-1)), \label{eq_kalmanfilter}  
\end{align}
where $K_f$ is the filter gain, $\bar{L}_{\mathrm{veh}}$ is average physical length of vehicles and $L_{\mathrm{d}}$ denotes sensor length. In the simulation, we set $K_f=0.1$ \cite{papamichail2010heuristic},. It should be pointed out that \eqref{eq_kalmanfilter} typically relies on internal detectors, e.g. induction loop sensors between detectors A and B in Fig.~\ref{fig_sumosetting}. However, we assume that such kind of sensors is not available and use detector A instead to provide the occupancy measurement $o_{\mathrm{A}}(t-1)$.  

We propose five scenarios for the purpose of comparison. The first and the second scenarios adopt the estimator \eqref{eq_openloopestimator}, but are given CV penetration rates $\alpha=0.25$ and $\alpha=0.5$, respectively. In scenario 3, the Kalman filter \eqref{eq_kalmanfilter} is considered.  We apply our robust filter in the last two scenarios, also given different CV penetration rates,  $\alpha=0.25$ and $\alpha=0.5$, respectively. For each scenario, we also add measurement noise between 15:00-17:00 in the simulation.

\begin{figure}[htbp]
    \centering
    \begin{subfigure}{\linewidth}
    \centering
    \includegraphics[width=\linewidth]{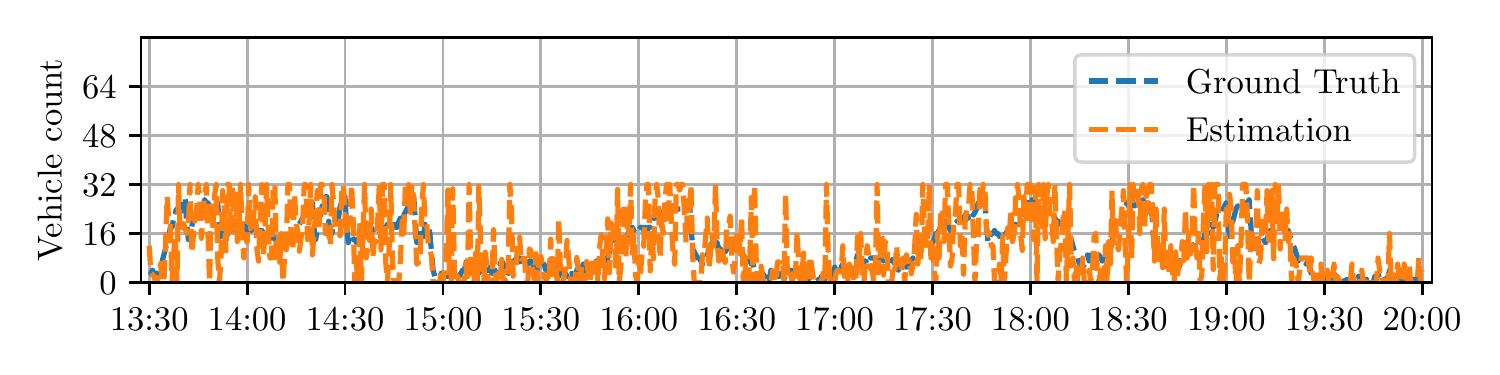}  
  \caption{Scenario 1: estimator \eqref{eq_openloopestimator} when $\alpha=0.25$.}
  \label{fig_scenario1}
\end{subfigure}

\begin{subfigure}{\linewidth}
  \centering
  \includegraphics[width=\linewidth]{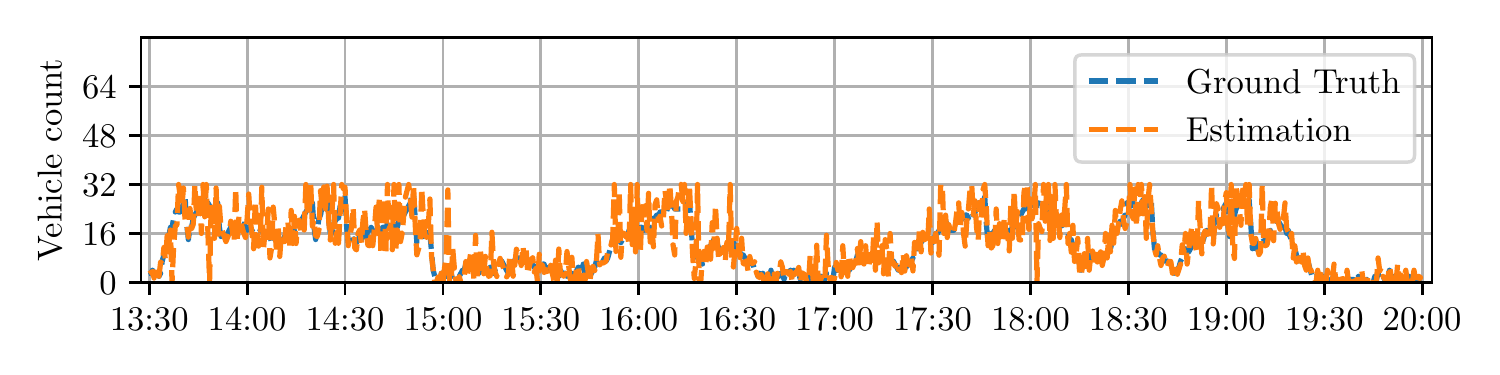}  
  \caption{Scenario 2: estimator \eqref{eq_openloopestimator} when $\alpha=0.50$.}
  \label{fig_scenario2}
\end{subfigure}

\begin{subfigure}{\linewidth}
  \centering
  \includegraphics[width=\linewidth]{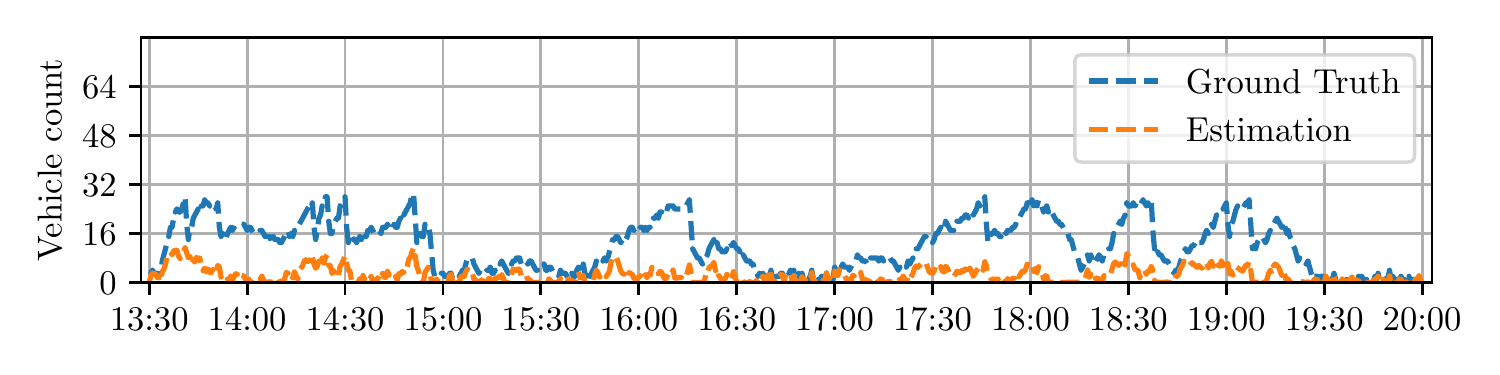}  
  \caption{Scenario 3: Kalman filter \eqref{eq_kalmanfilter} with gain $K_f=0.1$\cite{papamichail2010heuristic}.}
  \label{fig_scenario3}
\end{subfigure}

\begin{subfigure}{\linewidth}
  \centering
  \includegraphics[width=\linewidth]{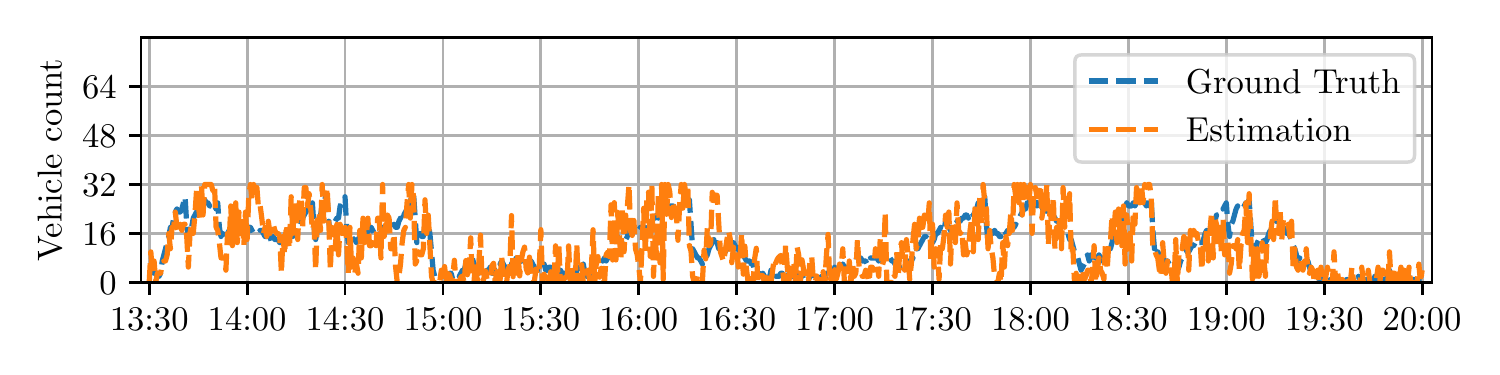}  
  \caption{Scenario 4: proposed robust filter when $\alpha=0.25$.}
  \label{fig_scenario4}
\end{subfigure}

\begin{subfigure}{\linewidth}
  \centering
  \includegraphics[width=\linewidth]{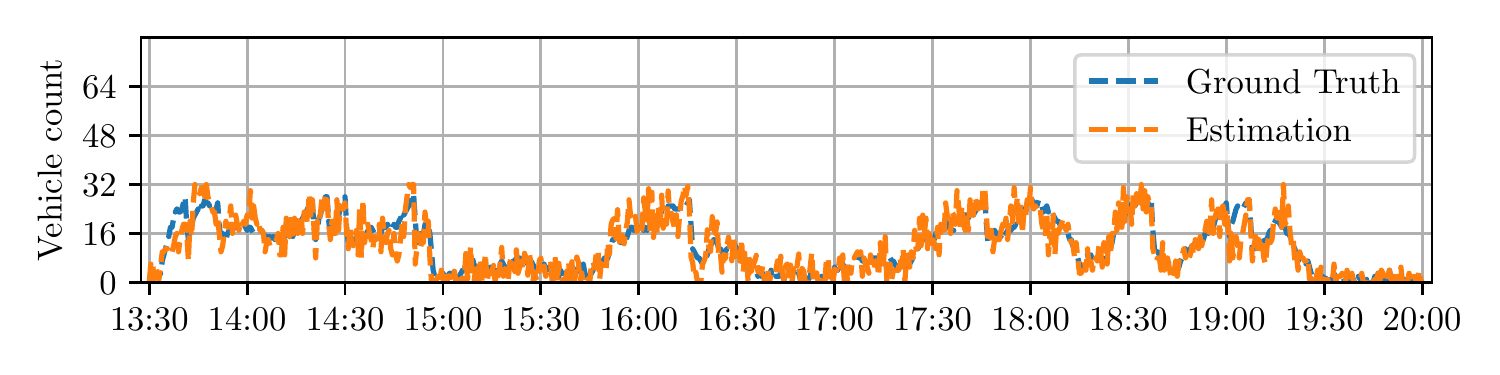}  
  \caption{Scenario 5: proposed robust filter when $\alpha=0.50$.}
  \label{fig_scenario5}
\end{subfigure}

    \caption{Comparison of different estimation methods.}
    \label{fig_comp}
\end{figure}

The results are shown in Fig.~\ref{fig_comp}. Clearly, the estimator \eqref{eq_openloopestimator} is sort of open-loop and thus produces noisier estimation than our robust filter does, given the same CV market penetration rate (see Fig.~\ref{fig_scenario1} vs. Fig.~\ref{fig_scenario4}, and Fig.~\ref{fig_scenario2} vs. Fig.~\ref{fig_scenario5}). Besides, it is not surprising to see that higher CV penetration rates benefit the estimation (see Fig.~\ref{fig_scenario1} vs. Fig.~\ref{fig_scenario2}, and Fig.~\ref{fig_scenario4} vs. Fig.~\ref{fig_scenario5}). When there are none of internal sensors, the Kalman filter \eqref{eq_kalmanfilter} could yield really poor performance. As indicated in Table~\ref{tab_comp}, it achieves relatively large $\sqrt{\hat{\mu}_1}$ and RMSE.
\begin{table}[htbp]
\centering
\footnotesize
\caption{Performance summary.}
\label{tab_comp}
\begin{tabular}{@{}ccc@{}}
\toprule
 & $\sqrt{\hat{\mu}_1}$  & RMSE \\ \midrule
Scenario 1 & 0.5767 & 8.490  \\
Scenario 2 & 0.3526 & 5.191  \\
Scenario 3 & 0.8205 & 12.079  \\
Scenario 4 & 0.4248 & 6.253 \\
Scenario 5 & 0.2701 & 3.976  \\
\bottomrule
\end{tabular}
\end{table}
%%%%%%%%%%%%%%%%%%%%%%%%%%%%%%%%%%%%%%%%%%%%%%%%%%%%%%%%%%%%%%%%%%%%%%%%%%%%%%%%
\section{Concluding remarks}
\label{sec_conclusion}

In this paper, we investigated ramp queue estimation that combines information from traditional infrastructure sensors and emerging CV data. We first presented a ramp queue model that takes measurement noise and uncertain CV penetration rates into account. Then we used the theory of extended $H_\infty$ filters to propose a robust filter for our ramp queue model and to show the performance guarantee of our filter (Theorem~\ref{thm_1}). It turns out that Theorem~\ref{thm_1} can be used to reveal how the uncertainty of penetration rates influences long-term estimation accuracy. Possible future directions include integrating our filters into localized/coordinated ramp control and analyzing their control performance.

\appendix
\subsection{Proof of Theorem~\ref{thm_1}}

\label{app_pf_thm1}

Consider the following Lyapunov function:
\begin{equation}
    V(e(t)) = e^\mathrm{T}(t)Pe(t).
\end{equation}
where $P\succ0$. We also define
\begin{equation}
    \xi(t) := \begin{bmatrix}
        e(t) & x(t) & w(t)
    \end{bmatrix}^{\mathrm{T}} 
\end{equation}
and
\begin{equation}
    W := \begin{bmatrix}
        A-LC & \Delta A(t) & D-LE
    \end{bmatrix}. \label{eq_W}
\end{equation}
It follows
\begin{align}
    \Delta V(e(t)):=&V(e(t+1)) - V(e(t)) \\
    =& \xi^\mathrm{T}(t)W^\mathrm{T}PW\xi(t) - e^{\mathrm{T}}Pe(t).
\end{align}
Then we obtain
\begin{align}
        &\Delta V(e(t)) + e^{\mathrm{T}}(t)e(t) - \mu_1 x^{\mathrm{T}}(t)x(t) - \mu_2 w^{\mathrm{T}}(t)w(t) \nonumber \\
        =& \xi^{\mathrm{T}}(t)(W^\mathrm{T}PW+U)\xi(t), \label{eq_1111}
\end{align}
where 
\begin{equation}
    U:=\begin{bmatrix}
        -P+I & 0 & 0 \\
        0 & -\mu_1 I & 0 \\ 
        0 & 0 & -\mu_2 I
    \end{bmatrix}. \label{eq_U}
\end{equation}
Summing both sides of \eqref{eq_1111} over $t=0,1,\cdots,T$ yields
\begin{align}
    &\sum_{t=0}^T||e(t)||_2^2 - \mu_1\sum_{t=0}^T||x(t)||_2^2 - \mu_2\sum_{t=0}^T||w(t)||_2^2 \nonumber \\
    =& -V(e(T)) + \sum_{t=0}^T \xi^{\mathrm{T}}(W^{\mathrm{T}}PW+U)\xi(t)  + V(e(0)). \nonumber 
\end{align}
Noting $V(e(T))>0$ and $\gamma(x(0),\hat{x}(0)):=V(e(0))$, we conclude that the filter satisfies \eqref{eq_criterion} if $W^\mathrm{T}PW+U\prec0$.

Next, we apply the Schur complement lemma:
\begin{lmm}[Schur complement lemma \cite{duan2013lmis}] \label{lmm_schur}
    Let $X$ be a symmetric matrix of real numbers given by 
    \begin{equation*}
        X = \begin{bmatrix}
            A & B \\ B^{\mathrm{T}} & C
        \end{bmatrix}.
    \end{equation*} 
    If $A$ is invertible, then $X\prec0$ if and only if 
    \begin{equation*}
        A\prec0,~ C - B^{\mathrm{T}}A^{-1}B \prec 0.
    \end{equation*}
\end{lmm}

Noting
\begin{equation*}
    W^{\mathrm{T}}PW + U = U -  W^{\mathrm{T}}P(-P)^{-1}PW,
\end{equation*}
we conclude by Lemma~\ref{lmm_schur} that $W^\mathrm{T}PW+U\prec0$ requires
\begin{equation}
    \begin{bmatrix}
        -P & PW \\
        W^{\mathrm{T}}P & U
    \end{bmatrix} \prec 0. \label{eq_111111}
\end{equation}
Recalling \eqref{eq_W} and \eqref{eq_U}, we obtain
\begin{align}
    &\begin{bmatrix}
        -P & PW \\
        W^{\mathrm{T}}P & U
    \end{bmatrix} = \tilde{P}F(t)\tilde{I} + \tilde{I}^{\mathrm{T}}F^\mathrm{T}(t)\tilde{P}^{\mathrm{T}} \nonumber \\
    &+\begin{bmatrix}
    -P & PA-RC & 0 & PD-RE \\
    * & -P+I & 0 & 0 \\
    * & * & -\mu_1 I & 0 \\
    * & * & * & -\mu_2 I
    \end{bmatrix} \label{eq_11111}
\end{align}
where
\begin{subequations}
    \begin{align}
            R:=&PL \\
            \tilde{P}:=& \begin{bmatrix}
        \Theta P & 0 & 0 & 0 
    \end{bmatrix}^\mathrm{T}, \\
            \tilde{I}:=& \begin{bmatrix}
        0 & 0 & I & 0 
    \end{bmatrix}, \\
    F(t):=& \Delta A(t) / \Theta.
    \end{align}
\end{subequations}

\begin{lmm}[Young's inequality \cite{duan2013lmis}] \label{lmm_young}
Let $X\in \mathbb{R}^{m\times n}$ and $Y\in\mathbb{R}^{n\times m}$. Then for any scalar $\kappa>0$ and any $F\in\mathbb{R}^{n\times n}$ such that $F^\mathrm{T}F\preceq I$, there holds
\begin{equation}
    XFY+Y^\mathrm{T}F^{\mathrm{T}}X^{\mathrm{T}} \preceq \kappa^{-1} XX^{\mathrm{T}} + \kappa Y^{\mathrm{T}}Y.
\end{equation}
\end{lmm}

Noting $F^{\mathrm{T}}(t)F(t)\preceq I$, we use Lemma~\ref{lmm_young} to obtain
\begin{equation}
    \tilde{P}F(t)\tilde{I} + 
    \tilde{I}^{\mathrm{T}}F^{\mathrm{T}}(t)\tilde{P}^{\mathrm{T}}
    \preceq \mu_3^{-1} \tilde{P}\tilde{P}^{\mathrm{T}}  + \mu_3 \tilde{I}^{\mathrm{T}}\tilde{I}. \label{eq_1111111}
\end{equation}

Combining \eqref{eq_11111} and \eqref{eq_1111111}, we find that if
\begin{align}
    &\begin{bmatrix}
    -P & PA-RC & 0 & PD-RE \\
    * & -P+I & 0 & 0 \\
    * & * & (\mu_3-\mu_1) I & 0 \\
    * & * & * & -\mu_2 I
    \end{bmatrix} \nonumber \\
    & + \mu_3^{-1}\tilde{P}\tilde{P}^{\mathrm{T}} \prec 0,    \label{eq_222}
\end{align}
then \eqref{eq_111111} holds, which indicates that the filter with the gain $L=P^{-1}R$ satisfies the criterion \eqref{eq_criterion}.

Finally, using Lemma~\ref{lmm_schur} again, we obtain \eqref{eq_222} if \eqref{eq_thm} holds. \qed

%%%%%%%%%%%%%%%%%%%%%%%%%%%%%%%%%%%%%%%%%%%%%%%%%%%%%%%%%%%%%%%%%%%%%%%%%%%%%%%%
%\section{ACKNOWLEDGMENTS}

%The authors gratefully acknowledge the contribution of National Research Organization and reviewers' comments.

%%%%%%%%%%%%%%%%%%%%%%%%%%%%%%%%%%%%%%%%%%%%%%%%%%%%%%%%%%%%%%%%%%%%%%%%%%%%%%%%

\bibliographystyle{IEEEtran}
\bibliography{Bibliography}

\end{document}